\documentclass[accepted,copyright,creativecommons,noncommercial]{eptcs}
\providecommand{\event}{FMAS 2022} 

\usepackage{iftex}

\ifpdf
  \usepackage{underscore}         
  \usepackage[T1]{fontenc}        
\else
  \usepackage{breakurl}           
\fi

\usepackage{graphicx}
\usepackage{pgfplots}
\usepackage{tikz}
\usetikzlibrary{positioning,calc,arrows,shapes,automata}
\usepackage[nolist,nohyperlinks]{acronym}
\usepackage{multirow}
\usepackage{array}
\usepackage{ragged2e}
\usepackage{makecell}
\usepackage{units}
\usepackage{color, colortbl}
\usepackage{eufrak}
\usepackage[ruled,vlined,linesnumbered]{algorithm2e}
\usepackage{caption}
\usepackage{subcaption}

\newcolumntype{P}[1]{>{\centering\arraybackslash}p{#1}}
\newcolumntype{L}[1]{>{\raggedright\arraybackslash}p{#1}}

\begin{acronym}
    \acro{ble}[BLE]{Bluetooth Low Energy}
    \acro{sul}[SUL]{System Under Learning}
    \acro{dfa}[DFA]{Deterministic Finite Automaton}
    \acro{mat}[MAT]{Minimally Adequate Teacher}
    \acro{pta}[PTA]{Prefix Tree Acceptor}
    \acro{rpni}[RPNI]{Regular Positive Negative Inference}
    \acro{mqtt}[MQTT]{Message Queuing Telemetry Transport}
    \acro{iot}[IoT]{Internet of Things}
    \acro{cps}[CPS]{cyber-physical system}
\end{acronym}

\title{Active vs.~Passive: A Comparison of \\ Automata Learning Paradigms for Network Protocols}
\author{Bernhard K.~Aichernig\textsuperscript{1}
\institute{\textsuperscript{1}Institute of Software Technology\\Graz University of Technology\\Graz, Austria}
\email{aichernig@ist.tugraz.at}
\and
Edi Muškardin\textsuperscript{1,2}
\institute{\textsuperscript{2}Silicon Austria Labs\\TU Graz - SAL DES Lab\\ Graz, Austria}
\email{edi.muskardin@silicon-austria.com}
\and 
Andrea Pferscher\textsuperscript{1}
\institute{\textsuperscript{1}Institute of Software Technology\\Graz University of Technology\\Graz, Austria}
\email{andrea.pferscher@ist.tugraz.at}
}
\def\titlerunning{Active vs.~Passive: A Comparison of Automata Learning Paradigms for Network Protocols}
\def\authorrunning{B.~K.~Aichernig et al.}
\begin{document}
\maketitle

\begin{abstract}
Active automata learning became a popular tool for the behavioral analysis of communication protocols. The main advantage is that no manual modeling effort is required since a behavioral model is automatically inferred from a black-box system. However, several real-world applications of this technique show that the overhead for the establishment of an active interface might hamper the practical applicability. Our recent work on the active learning of Bluetooth Low Energy (BLE) protocol found that the active interaction creates a bottleneck during learning. Considering the automata learning toolset, passive learning techniques appear as a promising solution since they do not require an active interface to the system under learning. Instead, models are learned based on a given data set. In this paper, we evaluate passive learning for two network protocols: BLE and Message Queuing Telemetry Transport (MQTT). Our results show that passive techniques can correctly learn with less data than required by active learning. However, a general random data generation for passive learning is more expensive compared to the costs of active learning.
\end{abstract}


\section{Introduction} \label{sec:intro}

%

Behavioral models provide a useful tool for the analysis and verification of autonomous systems. However, the availability of such a model might be limited. Manually creating a model is tedious and even if a model exists, keeping it up-to-date presents an ongoing challenge. Cyber-physical systems might consist of many heterogeneous components, e.g.~sensors, which are accessible communication protocols. However, access to these components or to the protocol implementations might be restricted, especially in third-party components. Such an environment motivates the automatic generation of a behavioral model from components of an autonomous system.

Automata learning has successfully been applied to extract behavioral models out of black-box systems. Besides theoretical research and learning competitions \cite{DBLP:conf/icgi/LangPP98,DBLP:conf/fsmnlp/CombeHJ09,DBLP:conf/spin/JasperFSSMPHS17}, nowadays automata learning is successfully applied in practice. In recent years, research has focused on learning behavioral models of network protocols like BLE~\cite{DBLP:conf/fm/PferscherA21}, (D)TLS~\cite{DBLP:conf/uss/RuiterP15,DBLP:conf/uss/Fiterau-Brostean20}, MQTT~\cite{DBLP:conf/icst/TapplerAB17}, or TCP~\cite{DBLP:conf/cav/Fiterau-Brostean16}. All of them applied active automata learning to generate a behavioral model of an implementation of the tested network protocol. 


Active learning characterizes that the behavior is explored by active interaction with the \ac{sul}. Depending on the application area, the development of a learning setup that enables active querying might be a tedious process. For example, consider the learning of wireless network protocols: sent packets might get lost or arrive delayed. Such behavior could introduce non-determinism. Non-deterministic behavior might interfere with the desired learning algorithm, which usually requires the \ac{sul} to behave deterministically. To treat non-deterministic behavior, the implementation of fault-tolerant mechanisms is necessary. Such mechanisms \cite{DBLP:conf/fm/PferscherA21,DBLP:conf/uss/Fiterau-Brostean20} include, e.g., the repetition of queries or the introduction of a cache to determine the correct output for a repeated input.

In automata learning, we distinguish between two learning paradigms: active and passive learning. These two paradigms are different in the generation of the required data set from which the behavioral model is inferred. As outlined before, active algorithms interact with the \ac{sul} to generate the required learning data. In contrast, passive techniques use a given data set, e.g.~log files, to generate a behavioral model. Consequently, passive learning algorithms can only cover the behavior that the given data includes. However, passive learning techniques might be better suited if the development of a sufficient active learning setup is costly in terms of the effort required to enable fault tolerance. 

The previous work \cite{DBLP:conf/uss/RuiterP15,DBLP:conf/icst/TapplerAB17,DBLP:conf/cav/Fiterau-Brostean16,DBLP:conf/fm/PferscherA21} argues that active learning provides in-depth behavioral aspects of the \ac{sul}. However, none of the techniques have been compared with passively generated data. We raise the question if randomly generated data is good enough to cover important behavioral aspects of the \ac{sul}. Furthermore, we approach the reduction of the data set size required for correct passive learning.

In this paper, we will compare passive and active techniques for the learning of \ac{ble} devices and \ac{mqtt} protocol implementations. In previous work \cite{DBLP:conf/fm/PferscherA21}, we discussed the difficulties in the wireless learning of \ac{ble} protocol implementations on different devices via a handcrafted interface. One of the main challenges was to guarantee that the device was sufficiently reset before performing the next query. Furthermore, the results show that timing constraints might differ between the \ac{ble} devices. Tappler et al.~\cite{DBLP:conf/icst/TapplerAB17} faced similar challenges in learning the \ac{mqtt} protocol. They also required individual learning setups for different \ac{mqtt} server implementations. In passive learning, these challenges do not need to be handled during learning. Hence, we are interested to see if passive learning can achieve similar results to active learning.

The aim of this paper is not only to discuss whether an active or a passive learning technique is preferable. In addition, we want to approach the challenge of generating a minimal adequate sample for both learning paradigms. We denote the optimization of data by minimizing the sample. Based on these problems, we discuss the following three research questions:
\begin{description}
    \item[\textbf{(RQ 1)}] Can passive learning based on a random sample outperform active learning?
    \item[\textbf{(RQ 2)}] Does the considered active automata learning algorithm generate an optimal sample?
    \item[\textbf{(RQ 3)}] Can random sampling support active automata learning?
\end{description}

\emph{Structure.} The paper is structured as follows. In Sect.~\ref{sec:preliminaries}, we introduce background concepts of our performed evaluation. Section~\ref{sec:evaluation} introduces the methodology and presents the result of the performed case study. We show related work in Sect.~\ref{sec:related-work}. Finally, Sect.~\ref{sec:conclusion} concludes the paper with a summary, a discussion of the found results, and an outlook on future work.
\section{Background} \label{sec:preliminaries}

\subsection{Mealy machine}

Mealy machines represent a modeling formalism for the behavioral description of reactive systems. A Mealy machine is a finite state machine with transitions labeled with an input action and the corresponding observable output. We define a Mealy machine $\mathcal{M}$ as a 6-tuple $\langle Q, q_0, I, O, \delta, \lambda \rangle$ where  $Q$ is the finite state set,  $q_0$ is the initial state, $I$ is the finite set of inputs, $O$ is the finite set of outputs, $\delta: Q \times I \rightarrow Q$ is the state transition function, and  $\lambda: Q \times I \rightarrow O$ is the output function.

We assume that $\mathcal{M}$ is input-enabled. Hence, for every $q \in Q$ and $i \in I$, $\delta(q, i)$ and $\lambda(q, i)$ are defined. Furthermore, the definition of $\delta$ as a non-set-function implies that $\mathcal{M}$ is deterministic. Let $|S|$ denote the number of elements in a set $S$. We denote the size of $\mathcal{M}$ by the number of states, i.e. $|Q|$.

A sequence of input/output pairs is denoted as $s \in (I \times O)^*$, where $s^I \in I^*$ and $s^O \in O^*$ are the corresponding input and output sequences. We write $s \cdot s'$ for the concatenation of two sequences. Additionally, we write $\epsilon$ for the empty sequence and also lift a single element $e \in (I \times O)$ to sequences. Let $|s|$ express the number of input/output pairs in a sequence. We extend $\delta$ and $\lambda$ for sequences. The function $\delta^* : Q \times I^* \rightarrow Q$ returns the state reached after executing the given input sequence and $\lambda^* : Q \times I^* \rightarrow O^*$ returns the observed output sequence. An execution on $\mathcal{M}$ returns the output sequence of an input sequence starting from the initial state, i.e. $s^O = \lambda^*(q_0, s^I)$. A \emph{trace} is an input/output sequence of an execution on $\mathcal{M}$.
Let $T$ be the set of traces that can be generated from executions on a Mealy machine $\mathcal{M}$. 
We denote two Mealy machines $\mathcal{M}$, $\mathcal{M'}$ as equivalent if they generate the same set of traces $T$. 

\subsection{Automata Learning}
In automata learning, we infer a behavioral model from the system's data where this data could be given log files or actively queried. Generally, we assume that the \ac{sul} is a black box, where no insight into internal behavior is provided. Depending on the generation of data for learning, we distinguish between two paradigms: active and passive learning.

\subsubsection{Active Automata Learning}
Active automata learning actively queries the \ac{sul} to infer a behavioral model. For this, the algorithms require an interface that enables the querying of input sequences.

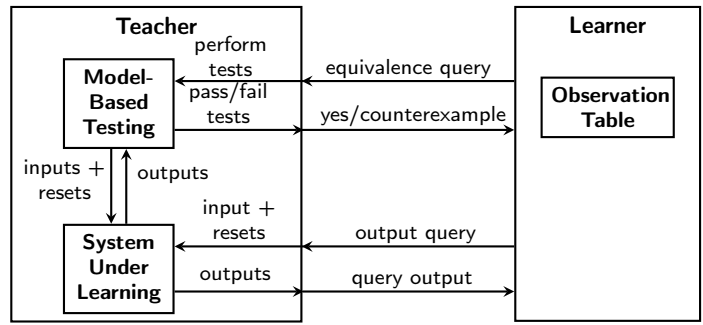
\begin{figure}[t]
    \centering
    	\begin{tikzpicture}[thick,font=\scriptsize\sffamily]
	\tikzstyle{bigbox} = [draw,font=\footnotesize\sffamily\bfseries,anchor=north,text centered,text depth = 3.7cm]
	\tikzstyle{tinybox} = [draw,font=\scriptsize\sffamily\bfseries ,minimum height = 0.75cm,  text width = 1.4cm,text centered]
	
	\node[bigbox,text width = 3.6cm] (teacher)  at (0,0){Teacher};
	\node[bigbox,text width = 2.2cm] (learner)  at (6,0){Learner};
	
	\node[tinybox, text width = 1.2cm] (ct) at  (-0.5,-1.3) {Model-Based Testing};
	\node[tinybox, text width = 1.2cm] (sut) at  (-0.5,-3.5) {System Under Learning};
	
	\node[tinybox, text width = 1.5cm] (obs-table) at  (6,-1.35) {Observation Table};
	
	\node[] (teacher-top-in) at (1.79,-1) {} ;
	\node[] (teacher-top-in-2) at (2.1,-1) {} ;
	\node[] (learner-top-out) at (4.9,-1) {} ;
	\node[] (ct-top) at (0.1,-1) {} ;
	
	\node[] (ct-bottom-out) at (0.1,-1.65) {} ;
	\node[] (ct-bottom-out-2) at (2.1,-1.65) {} ;
	\node[] (teacher-cex-out) at (1.79,-1.65) {} ;
	\node[] (learner-cex-in) at (4.9,-1.65) {} ;
	
	\node[] (perform-in) at (0,-1.2) {} ;
	
	\node[] (query-out) at (4.9,-3.2) {} ;
	\node[] (query-top-in-2) at (0.1,-3.2) {} ;
	\node[] (query-top-out-2) at (2.1,-3.2) {} ;
	\node[] (query-in) at (1.79,-3.2) {} ;
	\node[] (query-output-out-2) at (2.1,-3.8) {} ;
	\node[] (query-output-in-2) at (0.1,-3.8) {} ;
	\node[] (query-out-2) at (4.9,-3.8) {} ;
	\node[] (query-in-2) at (1.8,-3.8) {} ;
	
	\draw[->,>=stealth] (learner-top-out) edge (teacher-top-in) node [above left = -0.1cm and 0.25cm, text width = 6em, text centered] {equivalence query};
	\draw[->,>=stealth] (teacher-cex-out) edge (learner-cex-in) node [above  right= -0.1cm and 0.25cm, text width = 5em, text centered] {yes/counterexample};
	
	\draw[->,>=stealth] (query-out) edge (query-in) node [above left = -0.1cm and 0.25cm, text width = 5.5em, text centered] {output query};
	\draw[->,>=stealth] (query-in-2) edge (query-out-2) node [above right = -0.1cm and 0.4cm, text width = 5.5em, text centered] {query output};
	
	\draw[->,>=stealth] (query-top-out-2) edge (query-top-in-2) node [above left = -0.05cm and -0.2cm, text width = 5.5em, text centered] {input + \\ resets};
	\draw[->,>=stealth] (query-output-in-2) edge (query-output-out-2) node [above right = -0.05cm and -0.25cm, text width = 5.5em, text centered] {outputs};
	
	\draw[->,>=stealth,transform canvas={xshift=-0.1cm}] (ct) edge (sut) node [below left =0.6cm and -0.33cm, text width = 4.2em, text centered] {inputs + \\ resets};
	\draw[->,>=stealth,transform canvas={xshift=0.1cm}] (sut) edge (ct) node [above right =1.0cm and -0.33cm, text width = 4.2em, text centered] {outputs};
	
	\draw[->,>=stealth] (teacher-top-in-2) edge (ct-top) node [above left = -0.07cm and 0.2cm,  text width = 4em, text centered] {perform tests};
	\draw[->,>=stealth] (ct-bottom-out) edge (ct-bottom-out-2) node [above right = -0.05cm and 0.1cm, text width = 3.2em, text centered] {pass/fail tests};
	
	\end{tikzpicture}
    \caption{Angluin's \cite{DBLP:journals/iandc/Angluin87} adapted \ac{mat} framework. The adaptions relate to learning of reactive systems based on Smeenk et al.~\cite{DBLP:conf/icfem/SmeenkMVJ15} and Tappler et al.~\cite{DBLP:conf/icst/AichernigMP21}.}
    \label{fig:mat-framework}
\end{figure}

Many active learning algorithms build on Angluin's $L^*$ algorithm \cite{DBLP:journals/iandc/Angluin87}. In her seminal work, she introduced an algorithm that learns a \ac{dfa} representing an unknown regular language. For this, she introduced the \acf{mat} framework. Figure~\ref{fig:mat-framework} depicts a modified version of Angluin's \ac{mat} framework. The \ac{mat} framework distinguishes two members: the learner and the teacher. The teacher knows the regular language that is depicted as \ac{sul} and the learner wants to learn a behavioral model of the \ac{sul}.

The $L^*$ algorithm is an iterative procedure where each iteration includes two stages. According to the current stage, the learner asks two different kinds of questions: membership queries and equivalence queries. In the first stage, the learner aims to generate a hypothesis of the \ac{sul}. The hypothesis is a conjecture about the behavioral model of the \acf{sul}. For the creation of the hypothesis, the learner asks if a word is included in the regular language. The teacher answers these membership queries either by \emph{yes} or \emph{no}. The learner collects the answers of the teacher in an observation table. Based on the adequately filled observation table, the learner constructs a hypothesis. This hypothesis is then proposed to the teacher, who checks the equivalence to the \ac{sul}. The equivalence check between the provided hypothesis and \ac{sul} is considered the second stage of the learning algorithm. If they are equivalent, the teacher answers with \emph{yes} and the learning algorithm terminates by returning the proposed hypothesis. Otherwise, the teacher provides a counterexample to the conformance between the hypothesis and \ac{sul}, and the algorithm returns to the first stage. In the next iteration, the provided counterexample is then taken by the learner to retrieve new membership queries and to construct a new hypothesis. Hence, one iteration in active automata learning includes in the first stage the posing of several membership queries and in the second stage the equivalence check. We call one iteration \emph{learning round}. This iterative procedure is repeated until a conforming model is found. 

\begin{table}[t]
    \centering
    \scriptsize
    \caption{Observation table for Mealy machines from Shahbaz and Groz \cite{DBLP:conf/fm/ShahbazG09} with a modified $E$ set.}
    \begin{tabular}{|c|c|c|c|c|} \hline
         \multicolumn{2}{|c|}{} & \multicolumn{3}{c|}{$E$} \\  \cline{3-5}
         \multicolumn{2}{|c|}{} & $i_1$ & $i_2$ & $i_1 \cdot i_1$\\ \hline
         {$S$} & $\epsilon$ & $o_1$  & $o_1$ & $o_1 \cdot o_2$ \\ \hline
         \multirow{2}{*}{$S \cdot I$} & $ i_1$ & $o_2$  & $o_1$  & $o_2 \cdot o_1$\\  \cline{2-5}
         & $i_2$ & $o_1$  & $o_1$ & $o_1 \cdot o_2$ \\ \hline
    \end{tabular}
    \label{tab:observation-table}
\end{table}

The $L^*$ algorithm and the \ac{mat} framework have been extended for other modeling formalism. Shahbaz and Groz \cite{DBLP:conf/fm/ShahbazG09} inferred Mealy machines from reactive systems. For this, they modified the original \ac{mat} framework as follows: Instead of asking membership queries, output queries are posed. An output query is an input sequence that the teacher executes on the \ac{sul} and then returns the corresponding output sequence. Again, the learner stores the received output sequences in the observation table. For learning Mealy machines, the observation table can be described with a triplet $\langle S, E, T \rangle$, where $S$ and $E$ are non-empty sets of input sequences and $T : (S \cup S \cdot I) \times E \rightarrow O^*$ is a function with $I$ and $O$ as input and output set of the Mealy machine. $S$ is prefix-closed and $E$ is suffix-closed. According to Shahbaz and Groz \cite{DBLP:conf/fm/ShahbazG09}, $S$ is initialized with the empty sequence $\epsilon$ and $E$ with the inputs $I$. Table~\ref{tab:observation-table} shows an extended example of an observation table taken from Shahbaz and Groz \cite{DBLP:conf/fm/ShahbazG09}. The columns of an observation table are defined by $E$, and the rows by $S$ and $S \cup I$. The function $T$ accesses an output sequence stored in a table cell. 
A table row can be accessed via entries of $S \cup S \cdot I$ and is defined by the outputs of the corresponding values of $E$. The outputs in the cell $T(s, e)$ correspond to the outputs observed on the execution of $\lambda^*(\delta^*(q_0, s), e)$ with $s \in S \cup S \cdot I$ and $e \in E$. For example, the learner wants to query the output for a cell accessed by $T(i_2, i_1 \cdot i_1)$. For this, the learner asks the teacher the following output query: $i_2 \cdot i_1 \cdot i_1$, and the teacher responds with the query output $o_1 \cdot o_1 \cdot o_2$. The output sequence  $o_1 \cdot o_2$ of the $E$ set entry $i_1 \cdot i_1$ is then stored in the row of the observation table indexed by $i_2$. We refer to Shahbaz and Groz \cite{DBLP:conf/fm/ShahbazG09} for a detailed description of creating a Mealy machine from the observation table.

The application of automata learning is limited by the assumption of a perfect teacher that can determine the equivalence between a hypothesis and the \ac{sul}. In practice, we substitute the equivalence oracle with model-based testing techniques as shown in Fig.~\ref{fig:mat-framework}. Tretmans \cite{DBLP:conf/fortest/Tretmans08} introduces an implementation relation $\mathfrak{I}~\mathbf{imp}~\mathfrak{S}$ which defines that an implementation $\mathfrak{I}$ conforms to a specification $\mathfrak{S}$. In conformance testing, we want to find a test suite $T_\mathfrak{S}$ that can assess for every implementation $\mathfrak{I}$ if it implements a specification $\mathfrak{S}$. 
The aim of conformance testing in learning is to create a test suite that enables the assessment if a provided hypothesis $\mathfrak{H}$ conforms to implementation $\mathfrak{I}$ which represents our \ac{sul}. We say that the implementation $\mathfrak{I}$ passes a test sequence $t \in T_\mathfrak{H}$ $\mathfrak{I}~\mathbf{passes}~t$, if the output sequence generated from the execution on the learned model $\mathfrak{H}$ is equal to the obtained outputs sequence on $\mathfrak{I}$. Based on Tretmans implementation relation, we define conformance by the following equation. 
\begin{equation}
    \mathfrak{I}~\mathbf{imp}~\mathfrak{H} \quad \Longleftrightarrow \quad \forall t \in T_\mathfrak{H} : \mathfrak{I}~\mathbf{passes}~t. \label{eq:conformance-learning}
\end{equation}
The objective of conformance testing during learning is to find a test sequence that violates Eq.~\ref{eq:conformance-learning}. A found counterexample is then used to refine the hypothesis. In practice, the challenge of active automata learning is the selection of a conformance testing technique that is efficient in the number of required interactions with \ac{sul}, but also effective in the sense of finding counterexamples to the conformance between the hypothesis and the \ac{sul}.

\subsubsection{Passive Automata Learning}\label{sec:passive-automata-learning}
Passive automata learning infers an automaton from a given data set. Gold showed that the problem of inferring a \ac{dfa} with $k$ states from a given data set is \emph{NP-complete}~\cite{DBLP:journals/iandc/Gold78}.
State merging is one of the key technologies that is used in passive learning algorithms like in the \ac{rpni} \cite{OncinaG92,deLaHigueraRPNI} algorithm. \ac{rpni} takes a set of positive and negative traces. Positive traces include behavior that should be described by the learned automaton, whereas the behavior shown in negative traces must not be included. Based on the positive traces, \ac{rpni} builds a \ac{pta}. States of the \ac{pta} are then merged to create a finite automaton. A merge is valid if no negative trace can be generated by the merged automaton. Otherwise, if the merged automaton now includes negative examples, the merge is dismissed and a different merge is attempted.

\begin{figure}
    \centering
\begin{subfigure}[t]{0.38\textwidth}
\scalebox{0.75}{
    \begin{tikzpicture}[->,>=stealth',font=\sffamily,thick, node distance=2cm, every state/.style={thick}, initial text=$ $]
\node[state, initial, red] (q0) {$q_0$};
\node[state, above right = 0.5cm and 1cm of q0, blue] (q1) {$q_1$};
\node[state, below right = 0.5cm and 1cm of q0, blue] (q2) {$q_2$};
\node[state, right of=q1] (q3) {$q_3$};
\node[state, right of=q2] (q4) {$q_4$};
\node[state, right of=q3] (q5) {$q_5$};
\draw
(q0) edge[above] node[above left = 0cm and -0.2cm]{$i_1/o_1$} (q1)
(q0) edge[above] node[below left = 0cm and -0.2cm]{$i_2/o_1$} (q2)
(q1) edge[above] node{$i_1/o_2$} (q3)
(q2) edge[above] node[below]{$i_1/o_1$} (q4)
(q3) edge[above] node{$i_1/o_1$} (q5);
\end{tikzpicture}}
    \caption{Initial \ac{pta} generated from the two given traces in Sect.~\ref{sec:passive-automata-learning}. Blue nodes indicate candidates for merging with red nodes.}
    \label{fig:pta}
\end{subfigure}
\hfill
\begin{subfigure}[t]{0.37\textwidth}
\scalebox{0.75}{
    \begin{tikzpicture}[->,>=stealth',font=\sffamily,thick, node distance=2cm, every state/.style={thick}, initial text=$ $]
\node[state, initial, red] (q0) {$q_0$};
\node[state, above right = 0.5cm and 1cm of q0, red] (q1) {$q_1$};
\node[state, right of=q1, blue] (q3) {$q_3$};
\node[state, right of=q3] (q5) {$q_5$};
\draw
(q0) edge[above] node[above left = 0cm and -0.2cm]{$i_1/o_1$} (q1)
(q0) edge[loop below] node[]{$i_2/o_1$} (q2)
(q1) edge[above] node{$i_1/o_2$} (q3)
(q3) edge[above] node{$i_1/o_1$} (q5);
\end{tikzpicture}}
    \caption{Automaton after merging node $q_2$ into $q_0$ of the \ac{pta}.}
    \label{fig:pta-merged}
\end{subfigure}
\hfill
\begin{subfigure}[t]{0.2\textwidth}
\scalebox{0.75}{
    \begin{tikzpicture}[->,>=stealth',font=\sffamily,thick, node distance=3cm, every state/.style={thick}, initial text=$ $]
\node[state, initial, red] (q0) {$q_0$};
\node[state, right=1.25cm,red] (q1) {$q_1$};
\draw
(q0) edge[bend left, above] node[above]{$i_1/o_1$} (q1)
(q0) edge[loop below] node[]{$i_2/o_1$} (q2)
(q1) edge[bend left, below] node{$i_1/o_2$} (q0);
\end{tikzpicture}}
    \caption{Final merged automaton for the given traces.}
    \label{fig:pta-automaton}
\end{subfigure}
    \caption{Steps of the \ac{rpni} algorithm starting from the initial \ac{pta} to the final merged automaton.}
\end{figure}
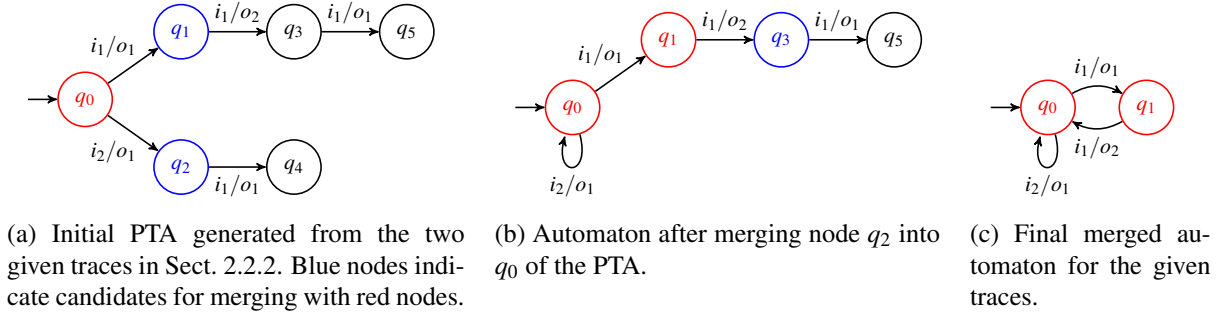

Variants of the \ac{rpni} algorithm can learn Mealy machines from input/output traces. For Mealy machines, states are merged iff outputs for a corresponding input are similar. For example, consider the following two traces:
\[
\begin{array}{lccccc}
        (1) & \texttt{($i_1$, $o_1$)} & \texttt{$\cdot$} & \texttt{($i_1$, $o_2$)} & \texttt{$\cdot$} & \texttt{($i_1$, $o_1$)} \\
        (2) & \texttt{($i_2$, $o_1$)} & \texttt{$\cdot$} & \texttt{($i_1$, $o_1$)} && 
        
\end{array}
\]
The \ac{pta} for these two traces is shown in Fig.~\ref{fig:pta}. To generate a more general behavioral model, \ac{rpni} merges states. The red states indicate the states that will be included in the final automaton and the blue states are currently candidates for merging with red states. For example, in the \ac{pta} presented in Fig.~\ref{fig:pta}, we first check if we can merge the blue state $q_1$ into the red state $q_0$. This merge is not possible since the input $i_1$ generates different outputs ($o_1$ vs.~$o_2$). However, $q_2$ can be merged into $q_0$. Figure~\ref{fig:pta-merged} shows the automaton after merging. In the next step, $q_3$ can be merged into $q_0$, which leads to the final automaton presented in Fig.~\ref{fig:pta-automaton}. Note that passive learning algorithms can only model behavior that is included in the given data set. For example, the learned automaton presented in Fig.~\ref{fig:pta-automaton} is not input enabled, since no trace was given that described the behavior of input $i_2$ in state $q_1$. 

\subsection{Bluetooth Low Energy}

\acf{ble} is a wireless communication protocol that has been introduced by Bluetooth Standard 4.0. Similar to the Bluetooth classic protocol, also known as Basic Rate, \ac{ble} is used for short-range peer-to-peer communication. However, \ac{ble} and Bluetooth classic are not compatible and depend on different implementations. The design of the \ac{ble} protocol also enables communication via Bluetooth for low-energy devices as they occur in the \ac{iot}. 

\begin{figure}[t]
    \centering
    \scalebox{0.8}{}{
    \begin{tikzpicture}[thick,font=\footnotesize\sffamily]
	
	\node[draw,text width = 1.5cm,minimum height=0.75cm,text centered,fill=gray!20!white] (central) {\textbf{central}};
	\node[draw,text width = 1.5cm,minimum height=0.75cm,text centered,right= 4cm of central,fill=gray!20!white] (peripheral) {\textbf{peripheral}};
	\node[text width = 1.5cm,minimum height=0.25cm,text centered,below= 6cm of central] (central-bot) {};
	\node[text width = 1.5cm,minimum height=0.25cm,text centered,below= 6cm of peripheral] (peripheral-bot) {};
	
	\node[text width = 0cm,minimum height=0.75cm,text centered,below right= 0.15cm and -0.88cm of peripheral] (ads-start) {};	
	\node[text width = 0cm,minimum height=0.75cm,text centered,below left= 0.15cm and 1.5cm of peripheral] (ads-end) {};
	
	\node[text width = 0cm,minimum height=0.75cm,text centered,below left= 0.65cm and -0.88cm of central] (scan-req-start) {};
	\node[text width = 0cm,minimum height=0.75cm,text centered,below right=  0.65cm and -0.88cm of peripheral] (scan-req-end) {};
	
	\node[text width = 0cm,minimum height=0.75cm,text centered,below = -0.2cm of scan-req-end] (scan-rsp-start) {};
	\node[text width = 0cm,minimum height=0.75cm,text centered,below = -0.2cm of scan-req-start] (scan-rsp-end) {};
	
	\node[text width = 0cm,minimum height=0.75cm,text centered,below = 0cm of scan-rsp-end] (con-req-start) {};
	\node[text width = 0cm,minimum height=0.75cm,text centered,below = 0cm of scan-rsp-start] (con-req-end) {};
	
	\node[text width = 0cm,minimum height=0.75cm,text centered,below = -0.2cm of con-req-end] (con-rsp-start) {};
	\node[text width = 0cm,minimum height=0.75cm,text centered,below = -0.2cm of con-req-start] (con-rsp-end) {};
	
	\node[text width = 0cm,minimum height=0.75cm,text centered,below = 0cm of con-rsp-end] (feature-req-start) {};
	\node[text width = 0cm,minimum height=0.75cm,text centered,below = 0cm of con-rsp-start] (feature-req-end) {};
	
	\node[text width = 0cm,minimum height=0.75cm,text centered,below = -0.2cm of feature-req-end] (feature-rsp-start) {};
	\node[text width = 0cm,minimum height=0.75cm,text centered,below = -0.2cm of feature-req-start] (feature-rsp-end) {};
	
	\node[text width = 0cm,minimum height=0.75cm,text centered,below = -0.1cm of feature-rsp-end] (pairing-req-start) {};
	\node[text width = 0cm,minimum height=0.75cm,text centered,below = -0.1cm of feature-rsp-start] (pairing-req-end) {};
	
	\node[text width = 0cm,minimum height=0.75cm,text centered,below = -0.25cm of pairing-req-end] (pairing-rsp-start) {};
	\node[text width = 0cm,minimum height=0.75cm,text centered,below = -0.25cm of pairing-req-start] (pairing-rsp-end) {};

	\draw
	(central.south) edge[->,>=latex] (central-bot.north)
	(peripheral.south) edge[->,>=latex] (peripheral-bot.north)
	
	(ads-start.west) edge[->,>=latex] (ads-end.east) node [above left, text width = 2cm] {advertisements}
	
	(scan-req-start.east) edge[->,>=latex] (scan-req-end.west) node [above right = -0.1cm and 2.25cm, text width = 1.5cm] {scan\_req}
	
	(scan-rsp-start.west) edge[->,>=latex,dashed] (scan-rsp-end.east) node [above left = -0.1cm and 1.75cm, text width = 1.5cm] {scan\_rsp}
	
	(con-req-start.east) edge[->,>=latex] (con-req-end.west) node [above right = -0.1cm and 1.75cm, text width = 3cm] {connection\_req}
	
	(con-rsp-start.west) edge[->,>=latex,dashed] (con-rsp-end.east) node [above left = -0.1cm and 0.75cm, text width = 3cm] {connection\_rsp}
	
	(feature-req-start.east) edge[<->,>=latex] (feature-req-end.west) node [above right = -0.1cm and 0.25cm, text width = 6cm] {\{length, feature, version, MTU\}\_req}
	
	(feature-rsp-start.west) edge[<->,>=latex,dashed] (feature-rsp-end.east) node [above left = -0.1cm and -0.75cm, text width = 6cm] {\{length, feature, version, MTU\}\_rsp}

	(pairing-req-start.east) edge[->,>=latex] (pairing-req-end.west) node [above right = -0.1cm and 2cm, text width = 6cm] {pairing\_req}
	
	(pairing-rsp-start.west) edge[->,>=latex,dashed] (pairing-rsp-end.east) node [above left = -0.1cm and -2.5cm, text width = 6cm] {pairing\_rsp}

	;

	\node[draw,dashed,text width = 1.5cm,minimum height=0.75cm,text centered,rounded corners,below left = 0.85cm and -0.5cm of central] (scanning) {scanning};
	\node[draw,dashed,text width = 1.5cm,minimum height=0.75cm,text centered,rounded corners,below = 0.75cm of scanning] (initiating) {initiating};
	\node[draw,dashed,text width = 1.5cm,minimum height=1.25cm,text centered,rounded corners,below = 0.75cm of initiating] (connectionM) {connection };
	
		\draw
	(scanning.south) edge[->,>=stealth'] (initiating.north)
	(initiating.south) edge[->,>=stealth'] (connectionM.north)
	;
	
	\node[draw,dashed,text width = 1.5cm,minimum height=0.75cm,text centered,rounded corners,below right = 0.15cm and -0.5cm of peripheral] (advertising) {advertising};
	\node[draw,dashed,text width = 1.5cm,minimum height=1.25cm,text centered,rounded corners,right = 6.6cm of connectionM] (connectionP) {connection};
	
	\draw(advertising.south) edge[->,>=stealth'] (connectionP.north);
	
\end{tikzpicture}}
    \vspace{-0.5cm}
    \caption{\ac{ble} sequence diagram for the establishment of a connection between two devices. The central device initiates the connection to the peripheral device. Note that after the initiation of the connection both devices could send requests that must be replied to by the other party. This figure is taken from \cite{DBLP:conf/fm/PferscherA21}.}
    \label{fig:ble-sequence-diagram}
\end{figure}
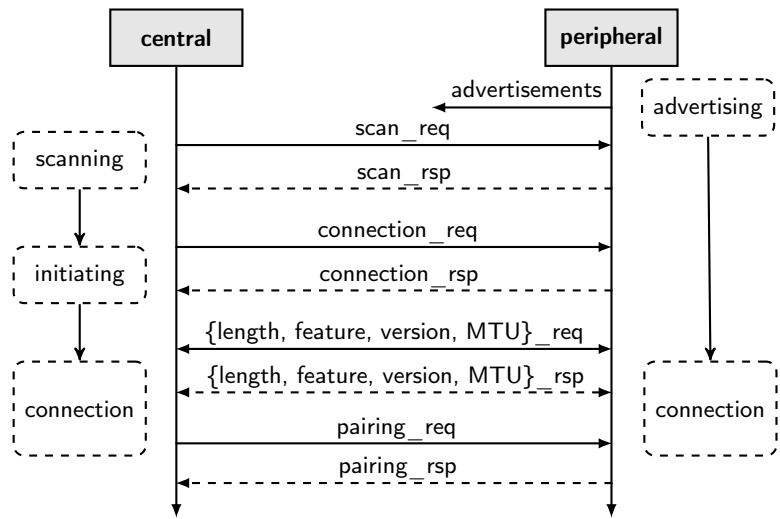

The Bluetooth standard~\cite{Bluetooth53} distinguishes between two communicating parties: the central and the peripheral device. For simplicity, we refer to the central device as \emph{central} and to the peripheral device as \emph{peripheral}. Figure~\ref{fig:ble-sequence-diagram} shows the sequence diagram of the connection procedure. The peripheral starts in the advertising state by sending advertisements to publicly announce that it is ready for a connection. The central device searches for advertisements by sending a scan request (\textsf{scan\_req}), to which the peripheral responds either by an advertisement or a dedicated scan response. Both answers are abstracted as \mbox{\textsf{scan\_rsp}}. Afterward, the central initiates the connection by sending a connection request (\textsf{connection\_req}), which is responded by a connection response (\textsf{connection\_rsp}). After the initiation of the connection, the central and the peripheral negotiate connection parameters. For example, the central can indicate the used Bluetooth version or request the maximum transmission unit (MTU). Both, the central and the peripheral, can send requests for specific parameters that should be valid for the current session. To further proceed to the pairing procedure these posed requests must be answered accordingly by the receiving party. After the negotiation phase, the pairing procedure can start. In the pairing procedure, keys for an encrypted communication are exchanged. 

\subsection{Message Queuing Telemetry Transport}

The \ac{mqtt} protocol \cite{mqttSpecification} is a publish/subscribe protocol. Due to its lightweight design, it is also a popular protocol for applications in the \ac{iot}. The \ac{mqtt} protocol distinguishes two members: the client and the broker. The broker is a central server unit that distributes messages from the clients. Clients connect to the broker and publish messages on a topic and the broker forwards then these published messages to clients who subscribed to the topic. In addition, some brokers offer advanced features like the definition of the client's will. The will of a client is a message assigned to a specific topic. The broker publishes the will of a client when the client disconnects. In the literature, we find several case studies \cite{DBLP:conf/icst/TapplerAB17,DBLP:conf/icst/AichernigMP21,DBLP:conf/pts/PferscherA20} which applied active automata learning to learn behavioral models of different \ac{mqtt}-broker implementations. The learned behavioral models revealed inconsistencies in the \ac{mqtt} specification. Note that the used modeling formalism requires the \ac{sul} to be deterministic. To simulate a deterministic behavior, the used learning setups required individual configurations since timed behavior led to non-deterministic observations. The observed non-determinism motivates the \ac{mqtt} case study for comparison to passive learning where non-determinism could be handled offline.
\section{Evaluation}\label{sec:evaluation}

In the following, we present our conducted case study and investigate the following research questions:

\begin{description}
    \item[\textbf{(RQ 1)}] Can passive learning based on a random sample outperform active learning?
    \item[\textbf{(RQ 2)}] Does the considered active automata learning algorithm generate an optimal sample?
    \item[\textbf{(RQ 3)}] Can random sampling support active automata learning?
\end{description}
Firstly, we introduce our applied methodology. For this, we present the used learning setup and the case study subjects. Furthermore, we explain the applied assessment technique of the gained learning results. Secondly, we discuss different sampling techniques that are used to generate data sets for passive learning. Finally, we present our gained results. The implemented framework is available \textbf{online}~\cite{supplementalMaterialPassiveEvaluation}.

\subsection{Methodology}\label{sec:methodolgy}

\noindent
\emph{Learning setup.} For automata learning, we used the learning framework \textsc{AALpy}~\cite{DBLP:conf/atva/MuskardinAPPT21} which is a Python library that implements many state-of-the-art automata learning algorithms. Originally, \textsc{AALpy} mainly focused on active automata learning algorithms. Beginning from the latest version (v.1.2.8), the tool also implements the passive learning algorithm \ac{rpni} for different modeling formalisms including Mealy machines. The latest version of \textsc{AALpy} is available on GitHub\footnote{\url{https://github.com/DES-Lab/AALpy}}.

For active learning, we use the $L^*$ algorithm for Mealy machines proposed by Shahbaz and Groz \cite{DBLP:conf/fm/ShahbazG09}.  For conformance testing during active learning, we applied a model-based testing technique that provides state coverage. Every state is accessed $n_\mathrm{walk}$ times via the shortest prefix and then a random input sequence of length $n_\mathrm{step}$ is executed. We set $n_\mathrm{walks} = 25$ and $n_\mathrm{step} = 30$. For counterexample processing, we applied the improved algorithm proposed by Rivest and Schapire \cite{DBLP:journals/iandc/RivestS93}. We used Rivest and Schapires $L^*$ algorithm since the performed benchmark for active automata learning conducted by Aichernig et al.~\cite{DBLP:conf/tap/AichernigTW20} concluded that this algorithm and the TTT algorithm~\cite{DBLP:conf/rv/IsbernerHS14} require approximately a similar number of queries and perform better compared to other active learning algorithms. 

For passive learning, we used a variant of the \ac{rpni} algorithm for Mealy machines implemented in \textsc{AALpy}. To test the conformance between two Mealy machines, we require input-enabledness. Since passive data might not be complete, the passively learned model might not be input-enabled. To overcome this problem, \textsc{AALpy} offers two strategies for conformance testing: self-looping transitions or transitions to a sink state. For our case study, we assume the transition to a sink state in the case of an undefined input. Hence, the execution of such an undefined input always leads to a counterexample to the conformance between the actively learned and passively learned model.

We performed all experiments presented in Table~\ref{tab:ble-active-results}, \ref{tab:mqtt-active-results}, \ref{tab:ble-passive-results}, and \ref{tab:mqtt-passive-results} on an Apple MacBook Pro 2019 with an Intel Quad-Core i5 with \unit[2.4]{GHz} and \unit[8]{GB} RAM. The experiments for the heatmaps (Fig.~\ref{fig:heatmap}) were conducted on a Dell Latitude 5410 with Intel Core i7-10610U with \unit[2.3]{GHz} and \unit[16]{GB} RAM.

\begin{figure}
    \centering
    \begin{tikzpicture}[>=stealth',font=\scriptsize\sffamily,thick,text width=0.4cm,scale=0.6, every node/.style={scale=0.8}] 
	\node[state,initial, align=center,initial text=,initial where=left,font=\small\sffamily] (q0) {$q_0$};
	\node[state, align=center,right = 2.5cm of q0,font=\small\sffamily] (q1) {$q_1$};
	\node[state, align=center,right = 3cm of q1,font=\small\sffamily] (q4) {$q_4$};
	\node[state,align=center,below = 1cm of $(q4)!0.5!(q1)$,font=\small\sffamily] (q2) {$q_2$};
	\node[state, align=center,font=\small\sffamily, above = 1.1cm of $(q4)!0.5!(q1)$] (q3) {$q_3$};
	
	\path[->] (q0.45) edge[bend left=45,looseness=1.1] node [above right=-0.05cm and -0.6cm,text width=2.5cm] {connect\_req\,/ SM\_RSP} (q1.150)
	(q1.180) edge[] node [below right=0.01cm and -0.75cm,text width=2.5cm] {scan\,/ADV length\_rsp\,/DATA} (q0.0)
	(q1.60) edge[bend left=0] node [above left=0cm and -0.25cm, align=right,text width=2cm] {version\_req\,/ VERSION\_IND} (q3.200)
	(q1.270) edge[bend right=0] node [below left=0cm and -0.15cm, align=right,text width=1.75cm] {mtu\_req\,/ MTU\_RSP} (q2.180)
	(q2.0) edge[bend right=40] node [below right=-0.5cm and 0.25cm, align=left,text width=2cm] {version\_req\,/ VERSION\_IND} (q4.270)
	(q3.0) edge[bend left=40] node [above left=0.15cm and -0.65cm, align=left,text width=1.75cm] {mtu\_req\,/ MTU\_RSP} (q4.90)
	(q3.270) edge[bend left=40] node [above right=0cm and -0.15cm, align=left,text width=1.75cm] {} (q1.40)
	(q2.100) edge[bend right=40] node [above right=-0.7cm and 0.5cm, align=left,text width=2cm] {connect\_req\,/ SM\_RSP} (q1.320)
	(q4.180) edge[bend right=0] node [above right = 0.15cm and -0.25cm, align=left,text width=2cm] {connect\_req\,/ SM\_RSP} (q1.0)
	(q3.150) edge[bend right=40] node [above right = 0.05cm and -0.25cm, align=left,text width=2.5cm] {scan\,/ADV length\_rsp\,/DATA} (q0.100)
	(q4.20) edge[bend right=100,looseness=1.3] node [above right = 0cm and -0.25cm, align=left,text width=2.5cm] {scan\,/ADV length\_rsp\,/DATA} (q0.140)
	(q2.240) edge[bend left=40] node [below left = -1cm and 1.5cm, align=left,text width=2.5cm, align=right] {scan\,/ADV length\_rsp\,/DATA} (q0.270)
	(q0) edge[loop,in=200,out=260,looseness=4] node [below=0cm,text width=1cm] {+/+} (q0)
	(q1) edge[loop,in=140,out=80,looseness=4] node [above=-0.1cm,text width=1cm] {+/+} (q1)
	(q2) edge[loop,in=270,out=340,looseness=4] node [right=0.1cm,text width=3cm] {mtu\_req\,/\,MTU\_ERR +/+} (q2)
	(q3) edge[loop,in=290,out=350,looseness=4] node [right=0cm,text width=1cm] {+/+} (q3)
	(q3) edge[loop,in=130,out=60,looseness=4] node [above right=-0.05cm and -1.25cm,text width=2.5cm,align=right] {version\_req\,/\,DATA} (q3)
	(q4) edge[loop,in=0,out=290,looseness=4] node [above right=-0.25cm and 0cm,text width=3cm] {\mbox{+/+} \mbox{version\_req\,/\,DATA} mtu\_req\,/\,MTU\_ERR} (q4)
	;
\end{tikzpicture}
    \caption{Model of the nRF52832. Input and output actions in the model are abbreviated or abstracted by the `+' character. The complete model and all other \ac{ble} models can be found online \cite{supplementalMaterialPassiveEvaluation}.}
    \label{fig:ble-model}
\end{figure}
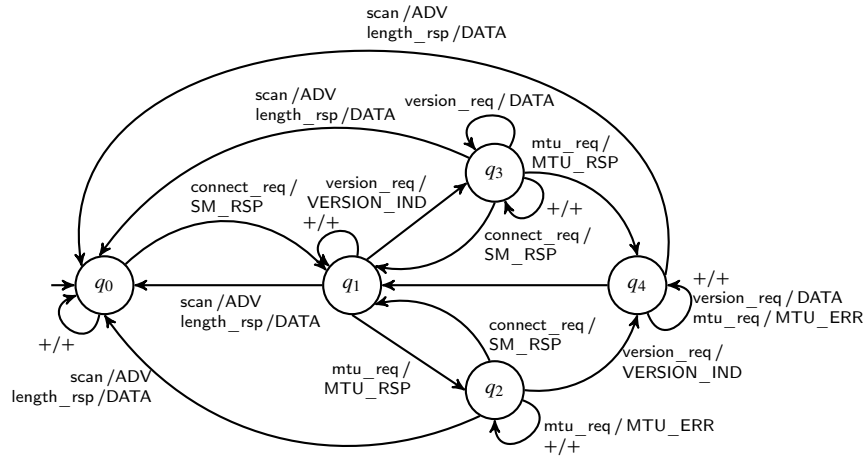

\noindent
\emph{Case study subject.} In our case study, we compare active and passive learning for two network protocols: \ac{ble} and \ac{mqtt}. For both protocols, we do not interact with the actual \ac{sul} but consider the already learned models from previous case studies \cite{DBLP:conf/nfm/PferscherA22,DBLP:conf/icst/TapplerAB17}. To generate output sequences, we simulate the given input sequence on the provided model, which also eases the reproducibility of our results. To maintain our black-box assumption, we do not access any further properties of the given models. Our previous work \cite{DBLP:conf/fm/PferscherA21} introduces the \ac{ble} case study which we extended in a follow-up work \cite{DBLP:conf/nfm/PferscherA22} by learning-based fuzzing. For this evaluation, we consider the \ac{ble} devices from our learning-based fuzzing case study \cite{DBLP:conf/nfm/PferscherA22}. Figure~\ref{fig:ble-model} depicts the model of  the \ac{ble} device \emph{nRF52832}. We indicate the initial state by a transition with an empty source. The input and output labels are abbreviated. The model of the nRF52832 describes that during an established connection the \mbox{\textsf{version\_req}} and \mbox{\textsf{mtu\_req}} lead to a different output after the first execution. 

For the \ac{mqtt} case study, we consider a subset of the learned automata presented by Tappler et al.~\cite{DBLP:conf/icst/TapplerAB17}. The \ac{mqtt} automata are available as a benchmark set on the \emph{Automata Wiki}~\cite{DBLP:conf/birthday/NeiderSVK97}. The considered subset comprises the models of five different \ac{mqtt} brokers. To evaluate larger models, the models include also the will-message functionality. All brokers interact with two clients, where one client connects with a will that the server retains. The other client can subscribe to the will's topic. 


\begin{table}[t]
\caption{The evaluation considers two case studies: \ac{ble} and \ac{mqtt}. All considered automata are available online \cite{supplementalMaterialPassiveEvaluation}.}
    \centering
    \scriptsize
    \begin{tabular}{l|c|c||l|c|c} 
         \multicolumn{3}{c||}{\textbf{BLE}} & \multicolumn{3}{c}{\textbf{MQTT}} \\ \hline
         \rowcolor{gray!20}
         \textit{SUL} & $|Q|$ & $|I|$ & \textit{SUL} & $|Q|$ & $|I|$\\ \hline
         CC2640 [excl.~feat] & 11 & 8 & ActiveMQ & 18 & 9 \\ \hline
         CC2640 [excl.~pair] & 6 & 8 & emqtt & 18 & 9 \\ \hline
         CC2650 & 5 & 9 & HBMQTT & 17 & 9 \\ \hline
         CC2652R1 & 4 & 7 & Mosquitto & 18 & 9\\ \hline
         CYBLE-416045-02 & 3 & 9 & VerneMQ & 17 & 9\\ \hline
         CYW43455 & 4 & 7 & & & \\ \hline
         nRF52832 & 5 & 9 & & & \\ 
    \end{tabular}
    \label{tab:case-study-subjects}
\end{table}

Table~\ref{tab:case-study-subjects} shows the investigated examples for the \ac{ble} and \ac{mqtt} case study including the number of states $|Q|$ and the input alphabet size $|I|$ of the ground-truth models. All considered models are available online\footnote{\url{https://github.com/apferscher/ble-learning-passive}}~\cite{supplementalMaterialPassiveEvaluation}.
Note that for the execution on real systems, the generated traces must have been checked in advance for non-deterministic behavior. However, for passive learning, this process has to be done only once and can be separated from the learning process. 

\emph{Result evaluation.} For measuring the correctness of the learned models, we need to compare the learned models with the ground truth. For this, we test the conformance between the \ac{sul} and the learned model. Let $\mathcal{M}$ be a Mealy machine that represents the behavior of the \ac{sul} and $\mathcal{M}_L$ be the learned Mealy machine. We define the following conformance relation. 
\begin{equation}
    \mathcal{M}_L~\mathbf{imp}~\mathcal{M} \quad \Longleftrightarrow \quad \forall t \in T : \mathcal{M}_L~\mathbf{passes}~t. \label{eq:conformance-eval}
\end{equation}
We consider $T$ as a finite set of traces generated from executions on $\mathcal{M}$. To generate these traces, we use two different model-based testing techniques: random and state coverage-based test generation. For both techniques, the size of the test set $T$ is set to $10\,000$ traces. 
The generation of random traces for conformance testing is equal to the one used to generate the random sample for passive learning as later described in Algorithm~\ref{alg:randDataGen}. We define the minimum trace length $n_\mathrm{min} = 3$ and the maximum length $n_\mathrm{max} = 32$. The second model-based testing technique is based on state coverage similar to the equivalence oracle used in active learning. For this, we generate a finite set of test sequences for each state, which include the access sequence to the state and a random word as a suffix. We set the test sequences per state to $\lceil \frac{10\,000}{|Q|} \rceil$, where $Q$ is the set of states of the \ac{sul}. The random word length that is executed in every state is set to $10$. We provide for every technique the percentage of passed test cases. In the remaining discussion, we refer to the conformance based on state coverage, if not stated otherwise.


\subsection{Data Generation}\label{sec:data-generation}

We compare active and passive automata learning based on the required data to learn the correct behavioral model of the \ac{sul}. We motivate this criterion from a theoretical and practical perspective. Theoretically, the conformance of a passively learned model to the behavior of the \ac{sul} depends on the given data set. The learned behavioral model can only cover model behavior that is included in the provided data set. Therefore, we approach the problem of finding an adequate data set that sufficiently covers the behavior of the \ac{sul}. 

From the practical perspective, previous work \cite{DBLP:conf/fm/PferscherA21,DBLP:conf/icst/TapplerAB17} showed that active automata learning could successfully create behavioral models of the actual black-box systems. However, the required active interaction and the error recovery mechanisms during learning hampered the applicability of active learning. The objective is to decrease the interaction with the \ac{sul} without missing behavioral information. 

To tackle the theoretical as well as the practical challenge, we first evaluate if randomly generated data sufficiently represents the behavior of the \ac{sul}. Secondly, we optimize the data generated during active automata learning under the premise that the \ac{sul}'s behavior is still adequately represented. Thirdly, we investigate if random sampling could support active learning.

\begin{algorithm}[t]
\caption{Generation of set of random input/output traces. \label{alg:randDataGen}}
\footnotesize 
\DontPrintSemicolon
\KwIn{data set size $n_\mathrm{data}$, minimum length $n_\mathrm{min}$, maximum length $n_\mathrm{max}$, \ac{sul} $\mathcal{M} = \langle Q, q_0, I, O, \delta, \lambda \rangle$}
\KwOut{$S_R \in (I \times O)^*$}
$S_R \longleftarrow [\;]$\;
\For{$i\leftarrow 1$ \KwTo $n_\mathrm{data}$}{
    $n_\mathrm{len} \longleftarrow randInt(n_\mathrm{min}, n_\mathrm{max})$ \; \label{line:randomLen}
    $s^I \longleftarrow [\;]$\;
    \For{$j\leftarrow 1$ \KwTo $n_\mathrm{len}$}{
        $s^I \longleftarrow s^I \cdot rand(I)$\; \label{line:randomInp}
    }
    $s^O \longleftarrow \lambda^*(q_0, s^I)$ \; \label{line:outputSeq}
    $S_R \longleftarrow S_R \cup trace(s^I, s^O)$\; \label{line:appendTrace}
}
\end{algorithm}

\noindent
\emph{Random Data Generation.} Algorithm~\ref{alg:randDataGen} describes the random data generation which depends on three parameters: data set size $n_\mathrm{data}$, the minimum length of the trace $n_\mathrm{min}$, and the maximum length of the trace $n_\mathrm{max}$. To generate traces, we execute input sequences on Mealy machine $\mathcal{M}$ of the \ac{sul}. The algorithm returns the set $S_R$ of generated traces. In Line~\ref{line:randomLen}, we select the input sequence length uniformly at random between the minimum and maximum length. Next, the algorithm selects uniformly at random an input from the input alphabet $I$ of the \ac{sul} (Line~\ref{line:randomInp}). In Line~\ref{line:outputSeq}, the input sequence is executed on the \ac{sul} to generate the corresponding output trace. Let $trace(s^I, s^O)$ be a function that creates a trace by a pairwise concatenation of inputs and outputs. In Line~\ref{line:appendTrace}, we append the generated trace to $S_R$. We repeat this procedure until we generate $n_\mathrm{data}$ traces.

As a baseline for our comparison, we run Rivest and Schapire's \cite{DBLP:journals/iandc/RivestS93} $L^*$ algorithm variant. Let $S_{L^*} \in (I \times O)^*$ be the set of traces that $L^*$ queries to correctly learn a Mealy machine of the \ac{sul}. Note $S_{L^*}$ also includes the conformance tests performed during equivalence checking. We include these traces since they are required to make a statement about conformance in a black-box setup.

For the evaluation with the random data generation, we considered the following four different setups:

\noindent
\textbf{Sample~1:} \texttt{Set of random sequences of size $|S_{L^*}|$ (rand $|S_{L^*}|$)}
This experiment setup evaluates the quality of a randomly generated set with approximately the same properties as the one for active learning.
The size of random sequences $n_\mathrm{data} = |S_{L^*}|$. We set the minimum $n_\mathrm{min}$ and maximum $n_\mathrm{max}$ accordingly to the mean length of traces for $S_{L^*}$. Let $\overline{n_{L^*}}$ be the mean length of the traces in $S_{L^*}$, then we set $n_\mathrm{min} = 1$ and $n_\mathrm{max} = \lfloor 2\overline{n_{L^*}} - 1 \rceil$\footnote{$\lfloor x \rceil$ denotes the nearest natural number to the real number $x$, where a distance of $0.5$ is associate to the higher natural number.}.

\noindent
\textbf{Sample~2:} \texttt{Set of random sequences of size $2*|S_{L^*}|$ (rand $2*|S_{L^*}|$)}
 This experiment setup evaluates if a randomly generated set with approximately the same properties but doubled in size as the one for active learning is sufficient. The other parameters, $n_\mathrm{min}$ and $n_\mathrm{max}$, are equal to \textbf{Sample~1}.
 
\noindent
\textbf{Sample~3:} \texttt{Set of longer random sequences (rand long)}
Given that we know the number of states of the minimal Mealy machine $\mathcal{M}$ representing the \ac{sul}, we evaluate if traces that correspond to the size of $\mathcal{M}$ are good enough. The randomly generated traces are at least $|Q|$ long, $n_\mathrm{min} = |Q|$, and have a maximum length of $n_\mathrm{max} = 2*|Q|$. The random set is of size $|S_{L^*}|$. 

\noindent
\textbf{Sample~4:} \texttt{Set of sufficient random samples (rand corr)}
The goal is to find a parameter setup that generates a random sample that sufficiently represents the \ac{sul}. For this, we evaluated how the trace length and the data set size influence the success of learning. Based on these two parameters we generate different random samples. 

\noindent
\emph{Optimized Data Generation.} Our second data set approaches the minimization of the data set generated by the $L^*$ algorithm. Due to the design of the used version of the $L^*$ algorithm, we assume that the generated data set is not minimal. The data set generated by the $L^*$  algorithm is not minimal due to two aspects: performed queries might be a prefix of other queries and a non-minimal characterization set. Firstly, considering the incremental state exploration of the $L^*$ algorithm, we assume that some traces are a prefix of other traces that are later queried in the $L^*$ algorithm. Since the \ac{rpni} algorithm relies on the construction of a \ac{pta}, traces that are a prefix of another trace do not provide any further knowledge for passive learning. Secondly, we can make a reduction of the required traces based on the characterization set of the learned Mealy machine. The characterization set is a set of input sequences that generate different output sequences for each state of the \ac{sul}. In the $L^*$ algorithm version proposed by Shahbaz and Groz \cite{DBLP:conf/fm/ShahbazG09}, introduced in Sect.~\ref{sec:preliminaries}, the $E$ set defines the characterization set and is initialized with the whole input alphabet. However, not necessarily all inputs are required to distinguish a state. Consider the observation table, Table~\ref{tab:observation-table}, presented in Sect.~\ref{sec:preliminaries}. In this table, the input $i_2$ does not provide any information that enables the distinction of states. Hence, $i_2$ is not part of the characterization set and is not required in set $E$. We, therefore, can remove all queries that are required to fill this column. To get an optimized data set, we calculate the characterization set based on the Mealy machine of the \ac{sul}. The reduced data set is then used as a base for passive learning and we expect the learned automaton to be equal to the Mealy machine of the \ac{sul}.

\emph{Cache in active learning.} Our third evaluation investigates if random data could support active automata learning. The learning library \textsc{AALpy} supports the caching of queries using a \ac{pta}. Before executing a query, the algorithm first checks if the query is already included in the \ac{pta}. If the \ac{pta} includes the query output, then the query is not performed on the \ac{sul}. This reduces the number of required interactions with the \ac{sul}. In our evaluation, we investigate the impact on the required \ac{sul} interactions with a cache initialized by random samples. For this, we initialize the cache of the $L^*$ algorithm with a random sample that has the same properties as the one initially required by $L^*$, i.e. \textbf{Sample~1}.






\subsection{Findings}

Table~\ref{tab:ble-active-results} and \ref{tab:mqtt-active-results} present the results of actively learning the \ac{ble} and \ac{mqtt} models using the $L^*$ algorithm. All learned models conform to the original models of the corresponding \ac{sul}. In addition, the tables show the minimized data size, where the minimization is performed as described in Sect.~\ref{sec:data-generation}. 

A comparison of both tables shows that the \ac{mqtt} example requires significantly more queries to learn the final model. We explain this increase based on two observations. Firstly, Table~\ref{tab:case-study-subjects} shows that the state space for the \ac{mqtt} examples is larger. Secondly, the \ac{mqtt} examples require more than one learning round. This circumstance substantially increases the sum of performed queries during active learning. All \ac{ble} models could be learned within one learning round. Hence, the conformance test during learning was only performed once and did not add any new behavior to the \ac{sul}. However, the active learning of the \ac{mqtt} examples requires at least three conformance testing rounds. Regarding the minimization of data, we see a significant decrease in the data that is actually required to learn correctly. This observation agrees with our assumption that the data set includes redundant and non-essential queries. 

\begin{table}[t]
    \centering
    \scriptsize
    \caption{Active learning results for the \ac{ble} case study. Every experiment was repeated five times. Every device could be learned within one learning round. Hence, the reported results were deterministic.}
    \begin{tabular}{|L{2.9cm}|P{1.2cm}|P{1.4cm}|c|P{1.4cm}|c|c|c|c|}
        \hline
         & \textbf{\makecell{CC2640\\excl.~feat.}} & \textbf{\makecell{CC2640\\excl.~pair.}} & \textbf{CC2650} & \textbf{CC2652R1} & \textbf{\makecell{CYBLE-\\416045-02}} & \textbf{CYW43455} & \textbf{nRF52832}\\ \hline
        Output queries & 704 & 384 & 405 & 196 & 243 & 784 & 405\\ \hline
        Output queries steps & 3136 & 1472 & 1458 & 588 & 729 & 3136 & 1458\\ \hline
        Conformance tests & 275 & 150 & 125 & 100 & 75 & 400 & 125 \\ \hline
        Conformance tests steps & 8925 & 4775 & 3950 & 3100 & 2325 & 12800 & 3950\\ \hline
        Sum queries & 979 & 534 & 530 & 296 & 318 & 1184 & 530\\ \hline
        Sum steps & 12061 & 6247 & 5408 & 3688 & 3054 & 15936 & 5408\\ \hline
        $\overline{n_{len}}$ & 12.32 & 11.70 & 10.20 & 12.46 & 9.60 & 13.46 & 10.20\\ \hline
        Optimized queries & 312 & 129 & 123 & 50 & 50 & 388 & 123\\ \hline
        Optimized $\overline{n_{len}}$  & 4.55 & 3.9 & 3.65 & 3.08 & 3.04 & 4.13 & 3.65\\ \hline
    \end{tabular}
    \label{tab:ble-active-results}
\end{table}
\begin{table}[t]
    \centering
    \scriptsize
    \caption{Active learning results for the \ac{mqtt} case study. Every experiment was repeated five times, numbers in brackets show the standard deviation. If no standard deviation is given, the deviation is zero.}
    \begin{tabular}{|L{2.9cm}|P{2cm}|P{2.2cm}|P{2cm}|P{2cm}|P{2cm}|}
        \hline
         & \makecell{\textbf{ActiveMQ}} & \makecell{\textbf{emqtt}} & \textbf{HBMQTT} & \textbf{Mosquitto} & \makecell{\textbf{VerneMQ}}\\ \hline
        \textit{Output queries} & 5890.40 (695.77) & 5900.60 (1834.84) & 4009.40 (533.83) & 4056.40 (846.08) & 4356.20 (976.87) \\ \hline
        Output queries steps & 55771.60 (14265.88) & 62134.80 (33276.46) & 33176.40 (8277.56) & 34335.80 (10979.61) & 33850.80 (12344.85) \\ \hline
        Conformance tests & 450 & 450 & 425 & 450 & 425\\ \hline
        Conformance tests steps & 14721.2 (70.33) & 14654.2 (24.69) & 13794 (16.41) & 14737 (88.12) & 13834.4 (87.33)\\ \hline
        Learning rounds & 5.20 (1.30) & 5.20 (1.30) & 3.60 (0.55) & 4.00 (1.00) & 5.20 (1.10) \\\hline
        Sum queries & 6340.40 (695.77) & 6350.60 (1834.84) & 4434.40 (533.83) & 4506.40 (846.08) & 4781.20 (976.87) \\ \hline
        Sum steps & 70492.80 (14208.67) & 76789.00 (33275.29) & 46970.40 (8282.37) & 49072.80 (10998.34) & 47685.20 (12364.07) \\ \hline
        $\overline{n_{len}}$ & 11.04 (1.18) & 11.77 (2.21) & 10.55 (0.86) & 10.85 (1.00) & 9.90 (0.66) \\ \hline
        Optimized queries & 1450 & 1450 & 959 & 1015 & 959 \\ \hline
        Optimized $\overline{n_{len}}$  & 6.26 & 6.26 & 5.91 & 6.05 & 6.05 \\ \hline
    \end{tabular}
    \label{tab:mqtt-active-results}
\end{table}



\begin{table}[t]
    \caption{Passive automata learning results for the \ac{ble} case study. Every experiment was repeated five times. The table gives the mean value and the corresponding standard deviation in brackets. For the number of correctly learned models, the sum over all five experiments is given.}
    \centering
    \scriptsize
    \begin{tabular}{|P{1cm}|L{1.7cm}|P{1.2cm}|P{1.2cm}|P{1.2cm}|P{1.2cm}|P{1.25cm}|P{1.3cm}|P{1.2cm}|c|}
        \hline
        & & \textbf{\makecell{CC2640\\excl.~feat.}} & \textbf{\makecell{CC2640\\excl.~pair.}} & \textbf{CC2650} & \textbf{CC2652R1} & \textbf{\makecell{CYBLE-\\416045-02}} & \textbf{CYW43455} & \textbf{nRF52832}\\ \hline
         \rowcolor{gray!20}
          & ${n_\mathrm{data}}$ & 979.00 & 534.00 & 530.00 & 296.00 & 318.00 & 1184.00 & 530.00 \\ \cline{2-9}
         \rowcolor{gray!20}
         & $\overline{n_\mathrm{len}}$ & 12.64 (0.25) & 11.46 (0.32) & 9.84 (0.30) & 12.69 (0.37) & \hspace{0.1cm}9.47 (0.23) & 13.54 (0.20) & 10.04 (0.24)\\ \cline{2-9}
         \rowcolor{gray!20}
         & Conformance (random) \% & 99.92 (0.01) & 99.93 (0.02) & 99.89 (0.06) & 100.00 (0.00) & 99.93 (0.06) & 100.00 (0.00) & 99.94 (0.03) \\ \cline{2-9}
         \rowcolor{gray!20}
         & Conformance (coverage) \% & 99.80 (0.02) & 99.86 (0.03) & 99.82 (0.08) & 100.00 (0.00) & 99.89 (0.08) & 100.00 (0.00) & 99.85 (0.09) \\ \cline{2-9}
         \rowcolor{gray!20}
         \multirow{-5}{*}{\makecell{\textbf{Sample 1} \\ \texttt{rand} \\ $|S_{L^*}|$ \\\vspace{0.25cm}}}& Correct model & 0 & 0 & 0 & 5 & 1 & 5 & 0\\ \Xhline{2\arrayrulewidth}
          & ${n_\mathrm{data}}$ & 1958.00 (0.00) & 1068.00 (0.00) & 1060.00 (0.00) & 592.00 (0.00) & 636.00 (0.00) & 2368.00 (0.00) & 1060.00 (0.00) \\ \cline{2-9}
         & $\overline{n_\mathrm{len}}$ & 12.49 (0.16) & 11.62 (0.15) & 10.02 (0.20) & 12.33 (0.24) & \hspace{0.1cm}9.61 (0.16) & 13.55 (0.10) & 10.10 (0.15) \\ \cline{2-9}
         & Conformance (random) \% & 99.97 (0.01) & 99.98 (0.03) & 99.98 (0.02) & 100.00 (0.00) & 100.00 (0.00) & 100.00 (0.00) & 99.97 (0.02) \\ \cline{2-9}
         & Conformance (coverage) \% & 99.92 (0.04) & 99.96 (0.06) & 99.96 (0.04) & 100.00 (0.00) & 100.00 (0.00) & 100.00 (0.00) & 99.93 (0.06) \\ \cline{2-9}
         \multirow{-5}{*}{\makecell{\textbf{Sample 2} \\ \texttt{rand}\\$2\,*$\,$|S_{L^*}|$ \\\vspace{0.25cm}}} & Correct model & 0 & 2 & 2 & 5 & 5 & 5 & 0\\ \Xhline{2\arrayrulewidth} \rowcolor{gray!20}
         & ${n_\mathrm{data}}$ & 979.00 & 534.00 & 530.00 & 296.00 & 318.00 & 1184.00 & 530.00\\ \cline{2-9}  \rowcolor{gray!20}
         & $\overline{n_\mathrm{len}}$ & 16.55 (0.05) & 9.03 (0.07) & 7.46 (0.09) & 5.99 (0.07) & \hspace{0.1cm}4.49 (0.10) & 24.08 (0.12) & 7.54 (0.07) \\ \cline{2-9} \rowcolor{gray!20}
         & Conformance (random) \% & 99.93 (0.04) & 99.87 (0.04) & 99.87 (0.06) & 99.81 (0.14) & 99.79 (0.10) & 100.00 (0.00) & 99.91 (0.03) \\ \cline{2-9} \rowcolor{gray!20}
         & Conformance (coverage) \% & 99.83 (0.09) & 99.76 (0.07) & 99.76 (0.08) & 99.86 (0.10) & 99.72 (0.11) & 100.00 (0.00) & 99.81 (0.04) \\ \cline{2-9} \rowcolor{gray!20}
         \multirow{-5}{*}{\makecell{\textbf{Sample 3} \\ \texttt{rand}\\\texttt{long}\\\vspace{0.25cm}}} & Correct model & 0 & 0 & 0 & 1 & 0 & 5 & 0 \\ \hline
    \end{tabular}
    \label{tab:ble-passive-results}
\end{table} 

\begin{table}[t]
    \caption{Passive automata learning results for the \ac{mqtt} case study. The data presentation is similar to Table~\ref{tab:ble-passive-results}. The values in the brackets indicate the standard deviation. }
    \centering
    \scriptsize
    \begin{tabular}{|P{1cm}|L{1.8cm}|P{1.2cm}|P{1.2cm}|P{1.2cm}|P{1.2cm}|P{1.2cm}|P{1.2cm}|}
        \hline
        & & \textbf{ActiveMQ} & \textbf{\makecell{emqtt}} & \textbf{HBMQTT} & \textbf{Mosquitto} & \textbf{\makecell{VerneMQ}}\\ \hline
         \rowcolor{gray!20}
          & ${n_\mathrm{data}}$ & 6340.00 & 6351.00 & 4434.00 & 4506.00 & 4781.00 \\ \cline{2-7}
         \rowcolor{gray!20}
         & $\overline{n_\mathrm{len}}$ & 11.00 (0.11) & 12.00 (0.05) & 10.46 (0.08) & 10.98 (0.06) & 10.07 (0.10)\\ \cline{2-7}
         \rowcolor{gray!20}
         & Conformance (random) \% & 99.97 (0.01) & 99.96 (0.01) & 99.85 (0.01) & 99.95 (0.00) & 99.95 (0.02) \\ \cline{2-7}
         \rowcolor{gray!20}
         & Conformance (coverage) \% & 99.95 (0.02) & 99.94 (0.02) & 99.79 (0.02) & 99.91 (0.01) & 99.92 (0.04) \\ \cline{2-7}
         \rowcolor{gray!20}
         \multirow{-5}{*}{\makecell{\textbf{Sample 1} \\ \texttt{rand} \\ $|S_{L^*}|$ \\\vspace{0.25cm}}}& Correct model & 0 & 0 & 0 & 0 & 0 \\ \Xhline{2\arrayrulewidth}
          & ${n_\mathrm{data}}$ & 12680.00 & 12702.00 & 8868.00 & 9012.00 & 9562.00 \\ \cline{2-7}
         & $\overline{n_\mathrm{len}}$ & 11.02 (0.06) & 11.99 (0.03) & 10.54 (0.03) & 11.00 (0.04) & 9.94 (0.05) \\ \cline{2-7}
         & Conformance (random) \% & 99.98 (0.01) & 99.97 (0.01) & 99.98 (0.01) & 99.98 (0.01) & 99.98 (0.01) \\ \cline{2-7}
         & Conformance (coverage) \% & 99.97 (0.01) & 99.97 (0.01) & 99.96 (0.02) & 99.96 (0.02) & 99.97 (0.02)  \\ \cline{2-7}
         \multirow{-5}{*}{\makecell{\textbf{Sample 2} \\ \texttt{rand}\\$2\,*$\,$|S_{L^*}|$ \\\vspace{0.25cm}}} & Correct model & 0 & 0 & 0 & 0 & 0\\ \Xhline{2\arrayrulewidth} \rowcolor{gray!20}
         & ${n_\mathrm{data}}$ & 6340.00 & 6351.00 & 4434.00 & 4506.00 & 4781.00 \\ \cline{2-7}  \rowcolor{gray!20}
         & $\overline{n_\mathrm{len}}$ & 26.99 (0.05) & 27.00 (0.02) & 25.52 (0.02) & 27.00 (0.15) & 25.45 (0.12)  \\ \cline{2-7} \rowcolor{gray!20}
         & Conformance (random) \% & 99.97 (0.01) & 99.97 (0.01) & 99.99 (0.01) & 99.95 (0.06) & 99.97 (0.02)  \\ \cline{2-7} \rowcolor{gray!20}
         & Conformance (coverage) \% & 99.96 (0.02) & 99.96 (0.02) & 99.97 (0.01) & 99.94 (0.06) & 99.97 (0.02)  \\ \cline{2-7} \rowcolor{gray!20}
         \multirow{-5}{*}{\makecell{\textbf{Sample 3} \\ \texttt{rand}\\\texttt{long}\\\vspace{0.25cm}}} & Correct model & 0 & 0 & 0 & 0 & 0  \\ \hline
    \end{tabular}
    \label{tab:mqtt-passive-results}
\end{table}

\begin{figure}
    \centering
    \begin{subfigure}[b]{0.5\textwidth}
    \includegraphics[width=1.15\linewidth]{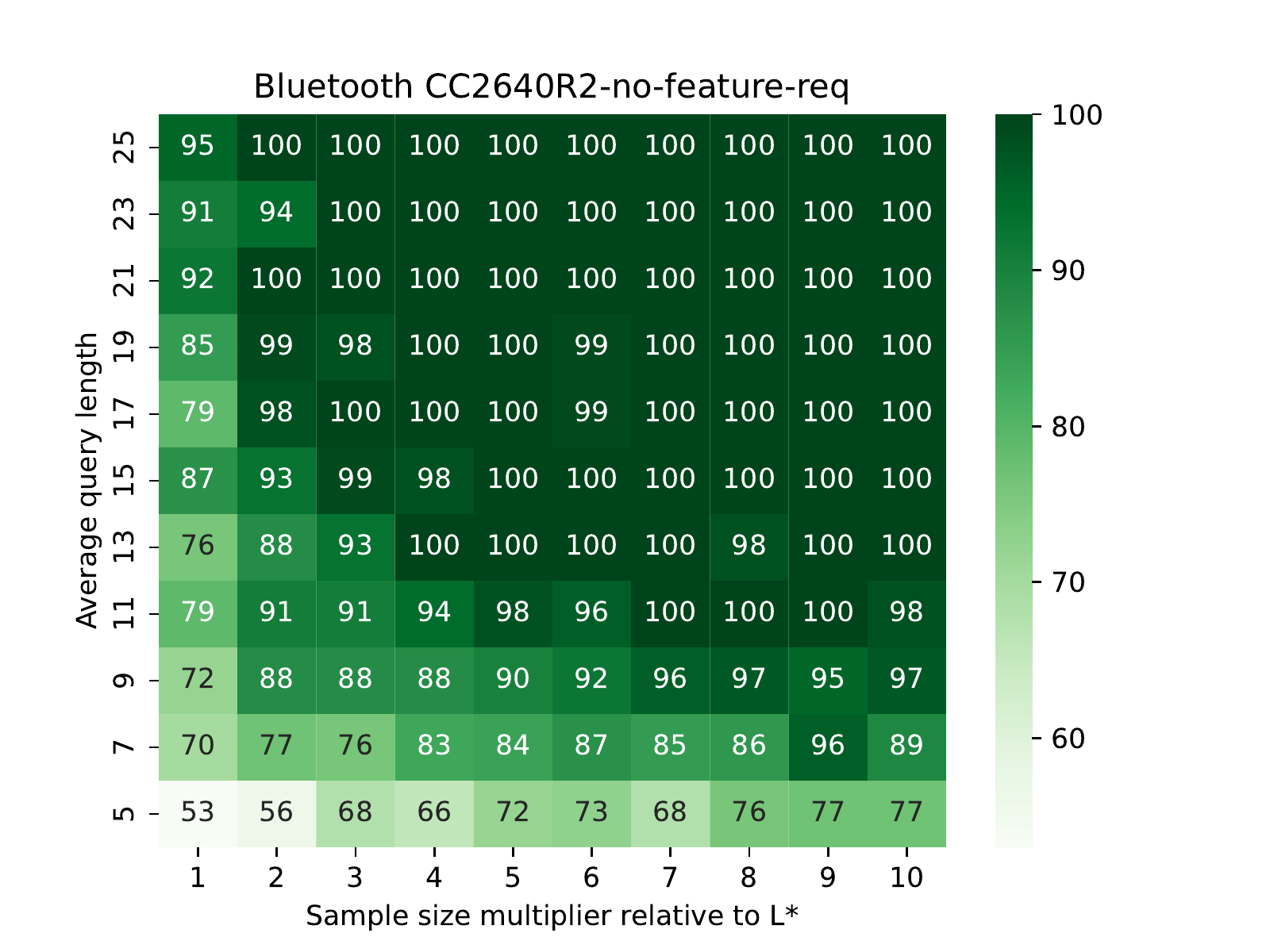}
    \label{fig:bluetooth-heatmap}
    \end{subfigure}%
    \begin{subfigure}[b]{0.5\textwidth}
    \includegraphics[width=1.15\linewidth]{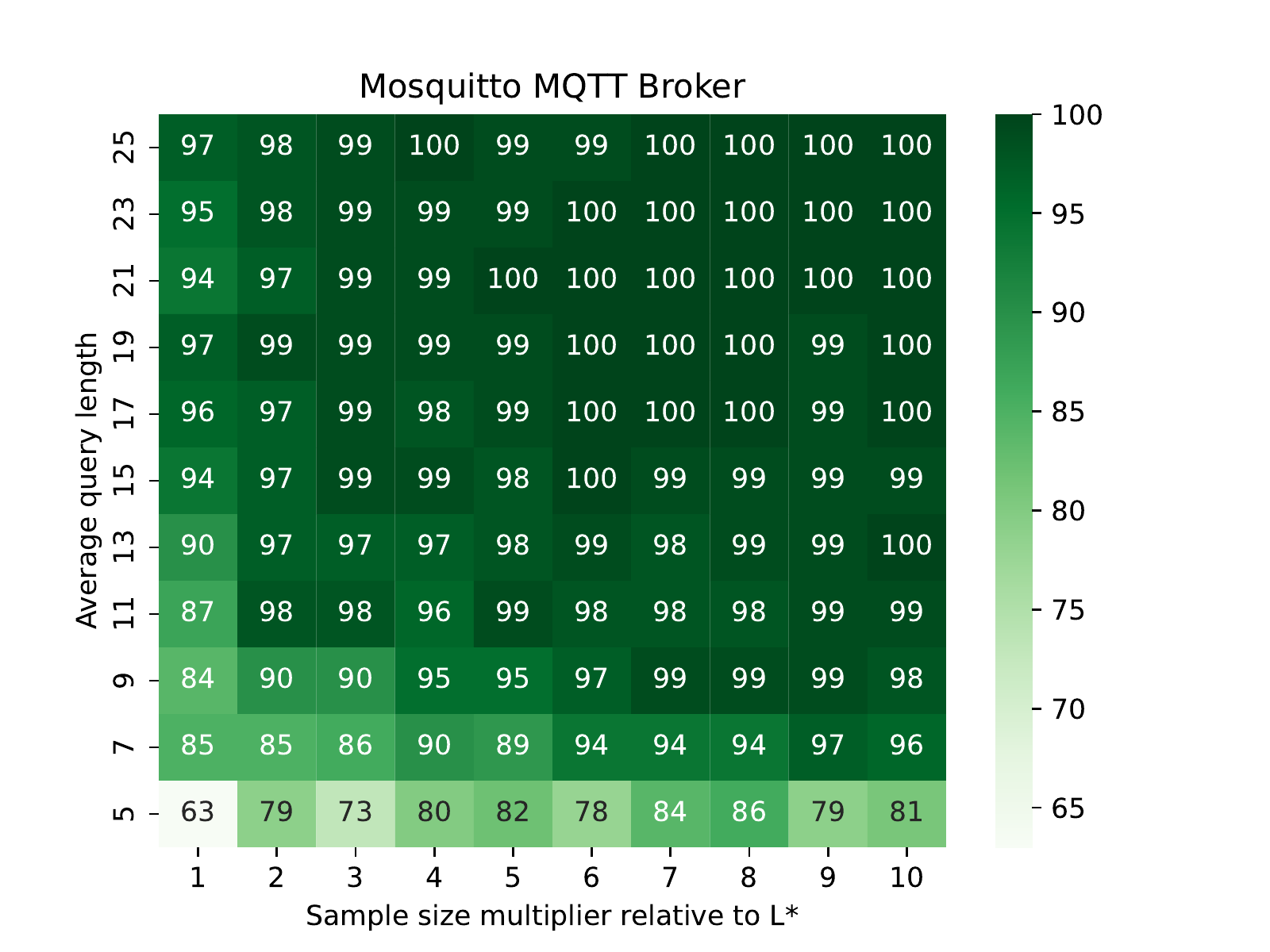}
    \label{fig:mqtt-heatmap}
    \end{subfigure}%
    \vspace*{-0.75cm}
    \caption{Each heatmap presents, for one example of the \ac{ble} and \ac{mqtt} case study, the influence of size and average trace length required to learn correctly.}
    \label{fig:heatmap}
\end{figure}

Table \ref{tab:ble-passive-results} and Table \ref{tab:mqtt-passive-results} present the evaluation results of passive learning considering the different randomly generated data sets \textbf{Sample~1-3}. For \textbf{Sample~4}, we created a heatmap which is presented in Fig.~\ref{fig:heatmap}. For each experiment setup, we state the size of the learning set $n_\mathrm{data}$, the mean length of the traces $\overline{n_\mathrm{len}}$, the conformance to the \ac{sul}, and the number of correctly learned models. Every experiment was repeated five times and we give the mean value of these five repetitions. We consider here one exception: the number of correctly learned models gives the absolute number and not the mean. For example, if we write ``1'' then one out of five learning repetitions learned the correct automaton.


In the following, we describe the obtained results based on our three research questions.

\textbf{(RQ 1)} \emph{Can passive learning based on a random sample outperform active learning?}
A comparison of Table~\ref{tab:ble-passive-results} and Table~\ref{tab:mqtt-passive-results} shows that the obtained results for two case studies provide a slightly different conclusion. For the \ac{ble} case study, in some cases, it was possible to learn the correct model with a random sample size equal to the active sample size. More precisely, for one \ac{ble} device, CYW43455, we always learn correctly with a random sample. We could also correctly learn for most experiments of the CC2652R1. The CYW43455 and CC2652R1 are two \acp{sul} with a smaller input alphabet size. We see that if the number of inputs increases, the chance to learn correctly decreases. Note that for the \ac{ble} case study the average length of \textbf{Sample~3} is shorter than the one of \textbf{Sample~1}. This is due to the fact that the generation of \textbf{Sample~3} is based on the size of the \ac{sul}.

For the \ac{mqtt} case study (Table~\ref{tab:mqtt-passive-results}), we never managed to learn correctly with the considered random samples. Even if active learning requires a large set of queries, passive learning seems to be no alternative for this case study. However, the passive learning conformance was always quite high with around $99\%$ conformance. A closer look at the passively learned automata revealed that the size of the learned automata was considerably larger than the size of the ground truth automata. For example, the average model size for HBMQTT based on \textbf{Sample~1} was $64.4$ instead of $17$. We assume that this increase in size results from sparse data. Since observations are missing, states could not be merged. 

For one \ac{ble} and \ac{mqtt} device, we evaluated how the average trace length and the number of samples influence the success of passive learning (\textbf{Sample~4)}. Figure~\ref{fig:heatmap} presents the results in two heatmaps. The x-axis indicates a factor by which the $L^*$ learning sample size $|S_{L^*}|$ is multiplied. The y-axis gives the mean length of the traces. The darker the green and the higher the number, the more the passively learned model conforms to the \ac{sul}. We observe that shorter traces are not sufficient to learn correctly, even if the sample size is large. However, to achieve $100\%$ conformance large random samples with long traces are required. Hence, considering that the interaction with the \ac{sul} should be decreased, passive learning does not provide an alternative.


\textbf{(RQ 2)} \emph{Does the considered active automata learning algorithm generate an optimal sample?}
Since passive learning does not improve the number of required interactions with the \ac{sul}, we investigate if there still is potential for optimization in active learning. To evaluate the potential of active learning, we generated an optimized learning sample set as described in Sect.~\ref{sec:data-generation}. With this optimized sample, passive learning could correctly learn the minimal model of the \ac{sul} for all examples. Table \ref{tab:ble-active-results} and \ref{tab:mqtt-active-results} show the same potential for improvement in both case studies. On average, the \ac{ble} active learning sample size could be reduced by $76\%$ and for \ac{mqtt} by $78\%$. For example, the \ac{ble} device CC2652R1 could be learned with a sample of size $50$ instead of $296$. Additionally, the average length can be significantly reduced. The reduction for \ac{ble} is on average $67\%$ and for \ac{mqtt} $43\%$. 

If we compare the experimental results between the random data sizes shown in Fig.~\ref{fig:heatmap} and the learning with the data set from $L^*$, we conclude that the generation of a sufficient data set via random techniques requires significantly more traces than the $L^*$ data set. For example, the heatmap on the right side representing the results of the \ac{mqtt} broker `Mosquitto' shows that even if the random data set size is $9$ times larger and the average trace length is approximately $9$ inputs longer than in the set required by $L^*$, passive learning still is not able to learn correctly. These results motivate the usage of active automata learning techniques. Based on the observation of Aichernig et al.~\cite{DBLP:conf/tap/AichernigTW20}, the choice of the learning algorithm and the corresponding conformance testing technique influences the required interaction with the \ac{sul}. Further improvements might be observable by the usage of other active learning algorithms, e.g.~TTT~\cite{DBLP:conf/rv/IsbernerHS14}, that avoid redundancy in the performed queries.

\textbf{(RQ 3)} \emph{Can random sampling support active automata learning?}

To reduce the redundancy of performed queries, we supported active automata learning by the usage of a cache. The motivation is to avoid query executions on the \ac{sul} since the observations can be retrieved from data already visible in the cache. For this, we initialized the cache with a random sample as it is generated in \textbf{Sample 1}. Hence, the size of the sample conforms to the number of queries required by the $L^*$ algorithm to learn correctly. In the performed experiments, we measure how many additional queries are required by the $L^*$ algorithm to generate a conforming model. 

Our obtained results show that a cache initialized by a random sample only covers the minority of the data that must be queried by active learning. On average, for the \ac{ble} case study, randomly generated data only represents $27\%$ of the total data required to learn correctly. As a result, additionally $63\%$ more queries had to be made by active learning. For MQTT, only $10.6\%$ of the performed queries were already present in the randomly initialized cache. 



\section{Related Work} \label{sec:related-work}

In the literature, the topic of model inference from network protocols is well studied. Several presented approaches used active automata learning to generate behavioral models of different protocols like 802.11 4-way handshake~\cite{DBLP:conf/esorics/StoneCR18}, BLE~\cite{DBLP:conf/fm/PferscherA21,DBLP:conf/nfm/PferscherA22}, (D)TLS~\cite{DBLP:conf/uss/RuiterP15,DBLP:conf/uss/Fiterau-Brostean20}, MQTT~\cite{DBLP:conf/icst/TapplerAB17}, QUIC~\cite{DBLP:journals/corr/abs-1903-04384}, SSH~\cite{DBLP:conf/spin/Fiterau-Brostean17}, or TCP~\cite{DBLP:conf/cav/Fiterau-Brostean16}. They motivate the usage of an active technique by considering it as a testing approach. Assuming that the system is input-enabled, every input is executed in every state. Therefore, input sequences are tested that might be rare during common usage.  This technique is also known as protocol state fuzzing since the execution of unusual input sequences might reveal unexpected behavior. 

There is also work that uses passive techniques. Comparetti et al.~\cite{DBLP:conf/sp/ComparettiWKK09} presented a general framework for passively learning communication protocols. Their evaluation includes protocols like malware bots, SMTP, or SIP. For learning, they used recorded sessions. Hence, their learned behavioral models only cover parts of the protocol that can be observed during normal user sessions. Other behavioral aspects have been excluded. Doupé et al.~\cite{DBLP:conf/uss/DoupeCKV12} generate behavioral models of web applications. Their proposed technique uses a web crawler to generate a set of traces that navigates through the web application. These crawled traces are then used to infer a model of the web application. Both passive techniques \cite{DBLP:conf/sp/ComparettiWKK09,DBLP:conf/uss/DoupeCKV12} use their inferred models to fuzz test the \ac{sul}.

The comparison of automata learning techniques mainly focuses on algorithms of the same paradigm. Initially, the Zulu challenge \cite{DBLP:conf/fsmnlp/CombeHJ09} addressed the problem of performing equivalence queries in active automata learning. The goal is to establish a conformance testing technique that can effectively test equivalence considering a limited testing budget. This challenge has now been replaced by the Rigorous Examination of Reactive Systems (RERS)~\cite{DBLP:conf/spin/JasperFSSMPHS17} challenge. RERS provides new benchmarks for evaluating learning techniques based on reactive systems. Aichernig et al.~\cite{DBLP:conf/tap/AichernigTW20} published a comprehensive evaluation of different combinations of active automata learning algorithms and conformance testing techniques. In their work, they conclude that the improved $L^*$ version from Rivest and Schapire \cite{DBLP:journals/iandc/RivestS93} or the TTT algorithm \cite{DBLP:conf/rv/IsbernerHS14} should be used for active learning. For conformance testing, they recommend the randomized version of the partial W-method~\cite{DBLP:journals/tse/FujiwaraBKAG91} or a mutation-based technique~\cite{DBLP:journals/jar/AichernigT19}.
For passive learning, Lang et al.~\cite{DBLP:conf/icgi/LangPP98} proposed a challenge to passively learn \acp{dfa} via state merging, including also the aspect of sparse data. The winning algorithm is based on evidence-driven state merging which was able to generalize better on sparse data. Lo et al.~\cite{DBLP:journals/jss/LoMS12} performed an empirical case study that evaluated different passive learning algorithms, e.g. kTail~\cite{DBLP:journals/tc/BiermannF72}, also considering sparse data.
A comparison of passive and active learning was performed by Aichernig et al.~\cite{DBLP:conf/nfm/AichernigPT20}. In their work, they compared a passive and an active variant of a search-based learning algorithm for timed automata. Their comparison shows that the active algorithm outperforms the passive variant in terms of required data to learn correctly.
\section{Conclusion}\label{sec:conclusion}

\noindent
\emph{Summary.} Active automata learning of network protocols has been an active research area in recent years. However, this technique might be limited in practice since a fault-tolerant learning setup that enables active interaction is required. This might hamper the application to autonomous systems. Passive automata learning, instead, infers a behavioral model from a given data set and does not require an interface that enables an active interaction with the \ac{sul}. Motivated by this, we compared active and passive learning by evaluating practical case studies. Our results show that we require a larger randomly generated sample for passive learning than the data set used by active automata learning. Furthermore, we showed that the data set generated by active learning can be optimized in terms of size. In summary, our results justify the costs of establishing an interface that allows an active interaction with the \ac{sul}.

\noindent
\emph{Discussion.} The primary goal for the inference of network protocols is to learn a conforming model with the lowest possible number of interactions. This goal is motivated by the fact that the interaction over a network to the \ac{sul} can be expensive in the sense that queries might be repeated since they were lost or arrived delayed. 
Our presented evaluation discusses if passive learning paradigms can overcome this problem of active interaction since they are based on a given data set. In practice, log files might represent such a data set. However, for our presented evaluation, we considered a randomly generated data set, since real-world logging data is assumed to be incomplete. For example, consider the \ac{ble} use case: logging data would probably not contain any parameter request that is not required for the establishment of a valid connection. 
The comparison with $L^*$ shows that even an active learning algorithm that queries a lot of redundant information is better than passive learning. However, our experiments on optimizing the data set of the $L^*$ algorithm indicate that there is room for improvements also on the side of active learning. We expect that other active algorithms, like TTT~\cite{DBLP:conf/rv/IsbernerHS14}, or other conformance testing techniques can help to improve active learning in the reduction of the required interaction with the \ac{sul}.

\noindent
\emph{Future Work.} For future work, it would be interesting to compare active and passive automata learning based on other practical case studies discussed in the literature. For example, to consider also additionally modeling formalisms like stochastic or timed systems, for which also active and passive learning algorithms exist. Additionally, a general comparison between several active and passive learning techniques on theoretical challenges, like RERS~\cite{DBLP:conf/spin/JasperFSSMPHS17}, could be useful. Furthermore, other learning algorithms should be considered to further optimize the required data set. A different direction would be the evaluation of a combination of passive and active learning as proposed by Walkinshaw et al.~\cite{DBLP:conf/fm/WalkinshawDG09}. Such an approach might unite the advantages of both learning paradigms since the proposed technique only requires an active interaction for conformance testing. Such a technique might reduce the required interaction with an \ac{sul} compared to an entirely active technique.

\noindent
{\it Acknowledgement.} This work was done in the TU Graz LEAD project ``Dependable Internet of Things in Adverse Environments'', the LearnTwins project funded by FFG (Österreichische Forschungs\-förderungs\-gesell\-schaft) under grant 880852, and the ``University SAL Labs'' initiative of Silicon Austria Labs (SAL) and its Austrian partner universities for applied
fundamental research for electronic based systems.

\bibliographystyle{eptcs}
\bibliography{references}
\end{document}